\newcommand{\rr}{{\mathbf r}}

\documentstyle[11pt,paspconf]{article}

\begin{document}

\title{Nonuniqueness and Structural Stability of Self-Consistent Models of
Elliptical Galaxies}
\author{Christos Siopis}
\affil{Department of Astronomy, University of Florida, P.O. Box 112055,
Gainesville, FL 32611-2055}

\vspace{8mm}

\noindent In recent years, high-resolution ground-based and HST
observations have shown that the space density profiles $\rho(r)$ near the
central regions of most elliptical galaxies exhibit power-law cusps, i.e.\
they scale as $\rho \sim r^{-\gamma}$ with $0 \leq \gamma \leq 2$. This
finding contradicts the previously widely held view that ellipticals have
extended central ``cores'' of near-constant density. Furthermore, evidence
has accumulated over the past two decades that the constant-density
surfaces of at least some ellipticals may actually be genuinely triaxial
ellipsoids rather than spheroids. This has spurred some interest in the
galactic community because theoretical and numerical work suggests that a
central density cusp or mass concentration embedded in a generic
nonaxisymmetric potential can lead to a significant expansion of the
phase-space regions where motion is chaotic.

In view of this evidence, Merritt and Fridman (1996) used Schwarzschild's
(1979) method to construct self-consistent models of a triaxial
generalization of Dehnen's spherical potential, which contains a central
density cusp. They found that self-consistency could be achieved only for
models with a weak ($\gamma = 1$) cusp, and then only when the chaotic
orbits populating the outermost regions of the model were allowed to be
not fully mixed (i.e., they were not required to sample the invariant
measure [Lichtenberg \& Lieberman, 1992]). Based on this and later work,
Merritt and collaborators have suggested that triaxiality may not in fact
be compatible with central density cusps, and that, as a consequence, most
elliptical galaxies may be (or may be evolving towards) axisymmetric
configurations, at least in the central regions.

The construction of a self-consistent equilibrium $\rho(\rr)$ via
Schwarzschild's numerical method is effected by assigning appropriate
weights to a large number of orbital templates, each evolved under the
influence of the gravitational potential generated by $\rho(\rr)$, so that
the weighted superposition of all the templates reproduces, at some level
of coarse graining, the initial $\rho(r)$. This is usually implemented via
some constrained optimization algorithm, such as linear or quadratic
programming. Owing to its conceptual simplicity and relative ease of
implementation, Schwarzschild's method has been used quite extensively for
the construction of stellar equilibria, especially when no known
analytical solutions to the collisionless Boltzmann equation exist or they
are difficult to compute.

In the present work (see also Siopis, 1999; Siopis \& Kandrup, 1999)
Schwarzschild's method was first used to construct self-consistent models
of a Plummer sphere and to calculate a number of velocity moments, which
were then compared with known analytical solutions (Dejonghe, 1986) to
assess the reliability of the numerical method. Subsequently, the method
was applied to the construction of triaxial Dehnen models with weak
($\gamma=1$) cusps. The principal moral derived from the extensive use of
Schwarzschild's method is that the importance of a good library of orbital
templates cannot be overemphasized. The initial conditions must be
selected carefully so as to provide a comprehensive coverage of phase
space. This usually means that one requires a good understanding of the
orbital structure of the system to be modeled. Failure to include enough
orbits, or an injudicious choice of initial conditions that misses
important families of orbits or violates the symmetries of the system, can
lead to unphysical results. Furthermore, special care should be taken to
ensure that each orbital template constitutes a truly time-independent
building block. This means that orbits should be integrated until they
uniformly cover their resonant tori (in the case of regular orbits) or
until they uniformly sample the invariant measure (in the case of chaotic
orbits).

Schwarzschild Plummer-sphere models were constructed both maximizing and
minimizing the number of near-radial (low angular momentum) orbits. The
resulting equilibria reproduce most of the structure that is present in
the velocity distributions computed analytically (Dejonghe, 1986). Where
agreement was less than satisfactory, the discrepancy could be traced to
inadequacies in the library of orbits. Since velocities were not
constrained explicitly in the construction of the models, these results
suggest that Schwarzschild's method can be used successfully to study the
degeneracy of the solutions.

Self-consistent models of the triaxial Dehnen potential could not be
constructed when the chaotic orbits at all energy levels were forced to be
completely mixed so as to yield time-independent building-blocks: only the
innermost 65\% of the mass could be mixed. In these inner regions, it was
possible to obtain alternative solutions that contain considerably
different numbers of chaotic orbits, yet yield (at least approximately)
the same mass density distribution. However, these solutions are not truly
time-independent, since the unmixed chaotic orbits in the outer regions,
which do not sample an invariant measure, will cause secular evolution.

Finally, some of the numerical equilibria were sampled to generate initial
conditions for N-body simulations to test the stability of the models.
Preliminary work showed that no catastrophic evolution takes place, but
there is a weak tendency for the configuration to become more nearly
axisymmetric over several dynamical times (Siopis et al., 1999). It is not
yet clear whether this tendency is real or a numerical artifact.

\end{document}